\documentclass[%
reprint,
superscriptaddress,
showpacs,preprintnumbers,
 amsmath,amssymb,
 aps,
prd,
]{revtex4-1}

\usepackage{graphicx}
\usepackage{dcolumn}
\usepackage{bm}
\usepackage{bbold}
\usepackage{amssymb,amsmath}
\usepackage{color}
\definecolor{refcol}{RGB}{0,0,205}
\usepackage[colorlinks,linkcolor=refcol,citecolor=refcol,urlcolor=refcol]{hyperref}

\usepackage{color}
\usepackage{amsfonts}
\usepackage{array}

\usepackage{placeins}
\usepackage{booktabs}
\usepackage{makecell}
\usepackage{subcaption}
\usepackage{floatrow}

\newcommand{\myhfill}{\hfill\textcolor{white}{.}}

\newcommand*{\no}{\noindent}
\newcommand*{\bea}{\begin{eqnarray}}
\newcommand*{\eea}{\end{eqnarray}}
\newcommand*{\be}{\begin{equation}}
\newcommand*{\ee}{\end{equation}}

\newcommand*{\pref}[1]{(\ref{#1})}

\newcommand*{\nn}{\nonumber}

\newcommand{\bma}{\begin{pmatrix}}
\newcommand{\ema}{\end{pmatrix}}

\graphicspath{{./figures/}{./}}

\begin{document}

\preprint{}

\title{Exploring the Tan contact term in Yang-Mills theory}
    


\author{Ouraman Hajizadeh}
\affiliation{Institute of Physics, NAWI Graz, University of Graz,Universit\"atsplatz 5, A-8010 Graz, Austria}

\author{Markus Q. Huber}
\affiliation{Institut f\"ur Theoretische Physik, Justus-Liebig-Universit\"at Giessen, Heinrich-Buff-Ring 16, 35392 Giessen, Germany}

\author{Axel Maas}
\affiliation{Institute of Physics, NAWI Graz, University of Graz,Universit\"atsplatz 5, A-8010 Graz, Austria}

\author{Jan M. Pawlowski}
\affiliation{Institut
 f\"{u}r Theoretische Physik, Universit\"{a}t Heidelberg,
 Philosophenweg 16, 69120 Heidelberg, Germany}
\affiliation{ExtreMe Matter Institute EMMI, GSI, Planckstr. 1, 64291
  Darmstadt, Germany}

\begin{abstract}

  Reliably computing the free energy in a gauge theory like QCD is a
  challenging and resource-demanding endeavor. As an alternative, we
  explore here the possibility to obtain the associated thermodynamic
  anomaly by exploiting its relation to the Tan contact. Optimally,
  this would reduce the determination of the free energy to a
  high-precision calculation of two-point correlators. We study this
  possibility using the lattice and functional methods and compare
  them to the expected behavior for the SU(2) Yang-Mills case.

\end{abstract}


\maketitle

\section{Introduction}

Thermodynamic observables including, e.g., density correlations are
among the most prominent observables that provide information about
the phase structure of heavy ion collisions. Their computations rely
on access to the bulk thermodynamic information of the system. This
information is encoded in the free energy, whose determination at
finite temperature and density is of prime interest. At large
densities functional approaches such as the Functional Renormalization
Group (FRG) and Dyson-Schwinger equations (DSEs) circumvent the
eminent sign problem that at present prevents lattice simulations in
this regime. However, while the part of correlations and thermodynamics
that comes from the matter fluctuations does not pose problems in
functional approaches, the access to the thermodynamics of gauge
fluctuations poses a formidable challenge beyond perturbation
theory. At its root it is related to the relevant momentum-scale
running of the thermodynamic potentials such as the free energy. Even
though by now functional methods have reached high quantitative
precision, such a computation still remains a demanding calculation in
terms of resources. This asks for computational approaches that
reduce the computational effort. 

Such an alternative may be provided by a computation in terms of the
Tan contact \cite{Tan:2008,Tan:2008gv}, for a discussion in Yang-Mills
theory see \cite{Fister:2011ym}. Essentially, it boils down to the
idea, rehearsed in section \ref{s:theo}, that the thermodynamic part
could be encoded in a simple way in the high-momentum behavior of the
two-point correlation functions. These are much simpler to determine
reliably. While the extracted part still needs some processing to
obtain the free energy, drastic features, e.\ g.\ phase transitions,
should manifest themselves already directly in the unprocessed
data. The aim of the present work is to explore exactly this
possibility.

To this end, we use the Landau-gauge gluon propagator of SU(2)
Yang-Mills theory at finite temperature. This theory undergoes a
second-order phase transition at a well established critical
temperature. Furthermore, its free energy is known quite well. It is
thus an ideal testbed for a new method. We use for this purpose the
gluon propagator as obtained using lattice methods, the functional
renormalization group, and Dyson-Schwinger equations. The results are
shown in section \ref{s:res}.

The Tan contact term is related in a straightforward way to the
thermodynamic anomaly. This is exploited in section \ref{s:anomaly},
where in a proof-of-principle style the anomaly is determined. There
it will also be discussed what further steps are required to make this
a quantitatively competitive approach.

In fact, the results show that interesting features of the
thermodynamics are already manifest for the unprocessed data. These
results are encouraging in that this is possible and may be a
promising future avenue, as is concluded in section \ref{s:con}.

\section{Setup}\label{s:theo}

The basic idea of how to transfer the Tan contact term formalism
\cite{Tan:2008, Tan:2008gv} from solid state physics and ultracold
atoms, e.g.\  \cite{Stewart:2010vo, Mukherjee:2019sr, Enss:2011va,
  Boettcher:2012dh, Braaten:2012ur}, to particle physics has been
outlined in \cite{Fister:2011ym}. It essentially boils down to that
the high-momentum behavior of a propagator $D(T,p_0,\vec p)$, i.\ e.\
at momenta $p\gg\Lambda_\text{YM}$, depending on both the temperature
$T$ and the momentum $p$, should behave essentially as
\begin{align}
  D(T,p_0^2,\vec
  p^2)=\frac{Z}{\frac{1}{ D_0(p_0^2+\vec p^2)}
  +C(T)D_0(p_0^2+\vec
  p^2)}\label{fitform}
\end{align}
where $C(T)$ is the Tan contact term, which is the only source of
temperature dependence. $D_0$ is the vacuum propagator and $Z$ is a
total normalisation of the propagator.  Here it is chosen such that
$Z\, D_0(\mu^2)=1/\mu^2$, i.e.\ we choose a temperature-independent
renormalization scheme.  By construction, the Tan contact term
satisfies $C(T=0)=0$ if the fit form \pref{fitform} describes the
propagator perfectly. Note also that the Tan contact in
\eqref{fitform} is not RG invariant, it runs with twice the anomalous
dimension of the propagator. An RG-invariant form is easily achieved by
multiplication with the wave function renormalization squared. As we
concentrate on the comparison between functional approaches and the
lattice, this is not important for us.

In the present work we consider the gluon propagator. As we are
interested in high energies, we set $D_0$ to be the one-loop resummed
propagator
\begin{align}
  D_0(p^2,\mu^2)=\frac{1}{p^2\left(1+\omega^2\ln\frac{p^2}{\mu^2}
  \right)^\frac{13}{22}}\,,\label{d0}
\end{align}
which entails $Z=1$. The quantity $\omega$ also involves the coupling
$g$. To accommodate for different renormalization prescriptions, we
fit $\omega^2$ to the zero-temperature propagator for the different
methods rather than to use some prescribed value. This approach
describes the gluon propagator above 2 GeV at zero temperature for all
methods at the 1-2\% level. In this regime also the propagator from
all methods coincide at this level of precision.

In addition, the thermal gluon propagator splits into a longitudinal
chromoelectric one and a transverse chromomagnetic one with respect to
the four-velocity of the heat-bath. Accordingly, we use \pref{fitform}
independently for both, thus computing a chromoelectric and
chromomagnetic Tan contact, $C_L(T)$ and
$C_\bot(T)$ respectively. To be in the asymptotic regime, we use only
data above $|\vec
p|>2$ GeV and the zeroth Matsubara frequency, though at these energies
the approximation $D(T,p_0^2,\vec p^2)\approx D(T,0,p_0^2+\vec
p^2)$ holds well anyway \cite{Fischer:2010fx}.

For the lattice case, we use the data from \cite{Maas:2011ez} with
some additional statistics and two additional lattice discretizations
at $T/T_c=0.9$ and $T/T_c=1.1$ with an $8\times 40^3$ lattice. For
the zero-temperature form \pref{d0} data from \cite{Maas:2014xma} are
used. This entails statistical errors on the fit parameters $Z$ and
$\omega$, which were propagated to the fit of $C(T)$. While
$\omega=0.82^{+0.04}_{-0.03}$ is essentially $\beta$-independent, $Z$ was
interpolated for different $\beta$ values by
$Z_01.50^{+0.06}_{-0.03}(\ln\beta)^{-1.11^{+0.01}_{-0.04}}$, where $Z_0$ is the
arbitrarily chosen renormalization prescription at zero temperature at
fixed $\mu=2$ GeV. In all cases fits where done along spatial
diagonals, which are least affected by discretization effects at high
momenta \cite{Maas:2014xma}.

For the FRG, we use the results from \cite{Cyrol:2017qkl}. The vacuum
results yield $Z=2.69Z_0$ and $\omega=0.795$ at $\mu=2$ GeV. The value
for $\omega$ agrees well with the lattice result.

We also extract the Tan contact term from DSE results.
Although they are obtained from a much simpler truncation than the
FRG results, the high momentum behavior
is determined sufficiently well to extract the relevant information
as shown below.
Details of the DSE calculations can be found in Appendix~\ref{sec:DSEs}. Their fit parameters are $Z=1.78Z_0$ and $\omega=0.752$, again in good agreement to the other methods.

The temperatures are taken from the respective works as well, i.e.\
we did not additionally try to fix any scales independently. This
yields agreement of the spatial-diagonal lattice data and the DSE and FRG
results from 2 GeV up to 12 GeV at the percent level and thus for
the whole range of relevant momenta in this work.

\section{Results}\label{s:res}
\begin{figure*}
	\ffigbox{
		\begin{subfloatrow}
			\ffigbox{\includegraphics[width=\columnwidth]{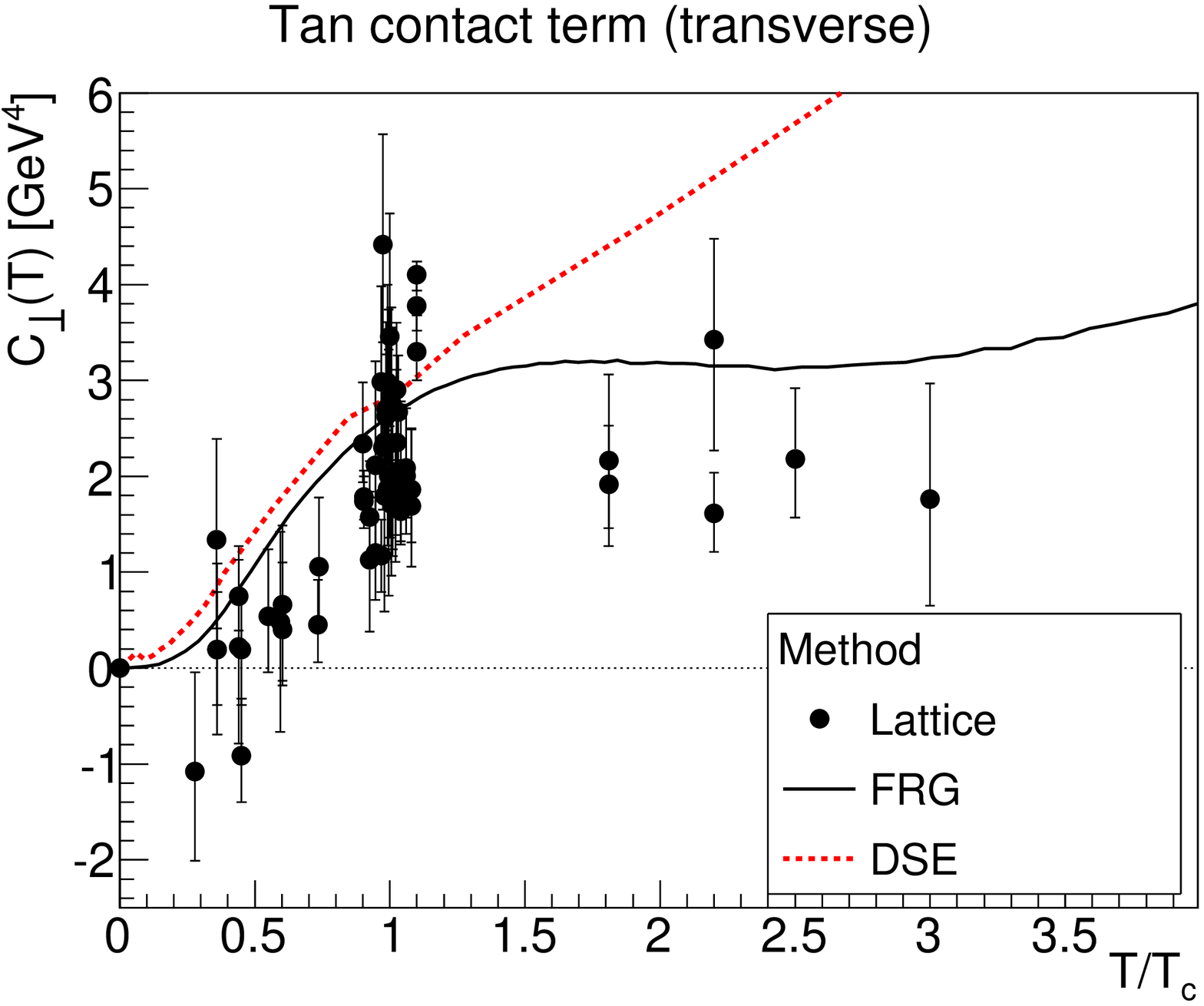}}{
                          \caption{Transverse $C_\bot(T)$ as a function of
                            the temperature.  \myhfill}
				\label{fig:ctt}
			}
			\ffigbox{\includegraphics[width=\columnwidth]{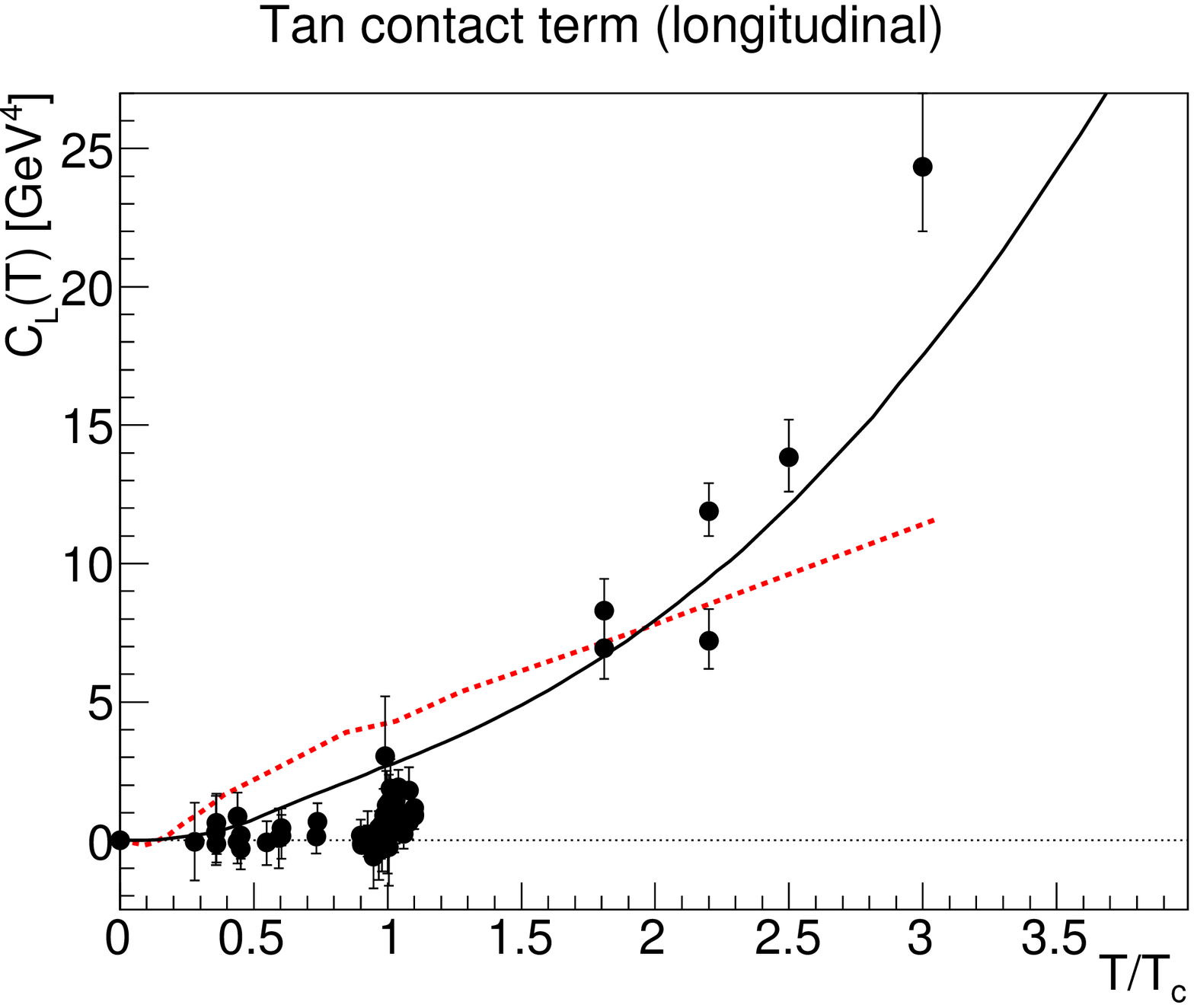}}{
                          \caption{Longitudinal $C_L(T)$ as a function
                            of temperature.\myhfill}
				\label{fig:ctl}
			} 
		\end{subfloatrow}
	}{
		\caption{
		Tan contact of Yang-Mills propagators from lattice, FRG \& DSE. 
		}
		\label{fig:TanYangMills}
	}
\end{figure*}

At finite temperature, we find that the fits work at the few percent
level well in all cases. However, results from the lattice for the two
temperatures $T/T_c=0.9$ and $T/T_c=1.1$ with ten times more
statistics reveal that \pref{fitform} is insufficient if at that level
of statistics a sub-percent fit is desired. Rather, $C(T)$ needs then
to be replaced by some extended form, e.\ g.\ $C(T)+p^2D(T)$. A
similar result is obtained in the FRG and DSE cases. However, for the
present purpose, and without a major effort for creating more
statistics for the lattice, we contend ourselves here with fits at the
2-3\% level, which at low statistics is also the statistical accuracy
of the lattice results, allowing for agreement within errors. Note
that for the continuum results the fit stops to work above roughly
$T/T_{c}\approx 5$. This is expected, as when $T$ becomes larger,
eventually screening effects will propagate to larger momenta which
are not included in the fit ansatz \pref{fitform}.

The results are shown in Fig.~\ref{fig:TanYangMills}. First of all, it is
visible that the general agreement between lattice and functional
methods is satisfactory, except for the DSE in the transverse case at high temperatures. Also, at this level of statistics no
statistically significant dependency on lattice parameters is
visible. Then, there are a number of visible trends which are
quite different for the transverse and the longitudinal Tan contact
term.

The probably most significant one is the difference between the
transverse one and the longitudinal one at high temperatures. The
transverse one starts to rise from essentially zero somewhere around
$t=T/T_c\approx 0.8$ for the lattice data, levels off shortly
after $t\gtrsim 1$, and stays constant up to $t\approx 3$. There is no
significant change happening at the phase transition. The functional
results switch on smoothly, but follow the same trend. However, above
$t\gtrsim 2.5$, the functional methods yield again a slow rise of the
Tan contact term.

The longitudinal one is quite different. Up to $t\approx 1$, the
lattice results are compatible with zero. There is a slight
systematic, though not statistically satisfactory trend to non-zero
values above $t=1$. However, at large temperatures the Tan contact
term rises quicker than quadratically with temperature. Except for the
smoothing of the transition, this behavior is also seen in the
functional results, this time with no particular impact at
$t\gtrsim 3$.

In comparison to the low-momentum behavior
\cite{Cyrol:2017qkl,Maas:2011ez}, this provides a consistent
picture. There, also the transverse propagator shows no substantial
impact of the phase transition, while the longitudinal one seems to do
so. At the same time, the impact at high temperatures is also stronger
for the longitudinal one.

This leads us to the following picture: The transverse sector carries
non-trivial thermodynamic behavior, which is sensitive to the
interactions which create a strongly-interacting phase above the
phase transition for a range of a few $T_c$. The bulk thermodynamics
is manifested in the longitudinal degrees of freedom, including both
the phase transition and the Stefan-Boltzmann trend at high
temperatures.

\section{The anomaly from the Tan contact term}\label{s:anomaly}
\begin{figure*}
	\ffigbox{
		\begin{subfloatrow}
			\ffigbox{\includegraphics[width=\columnwidth]{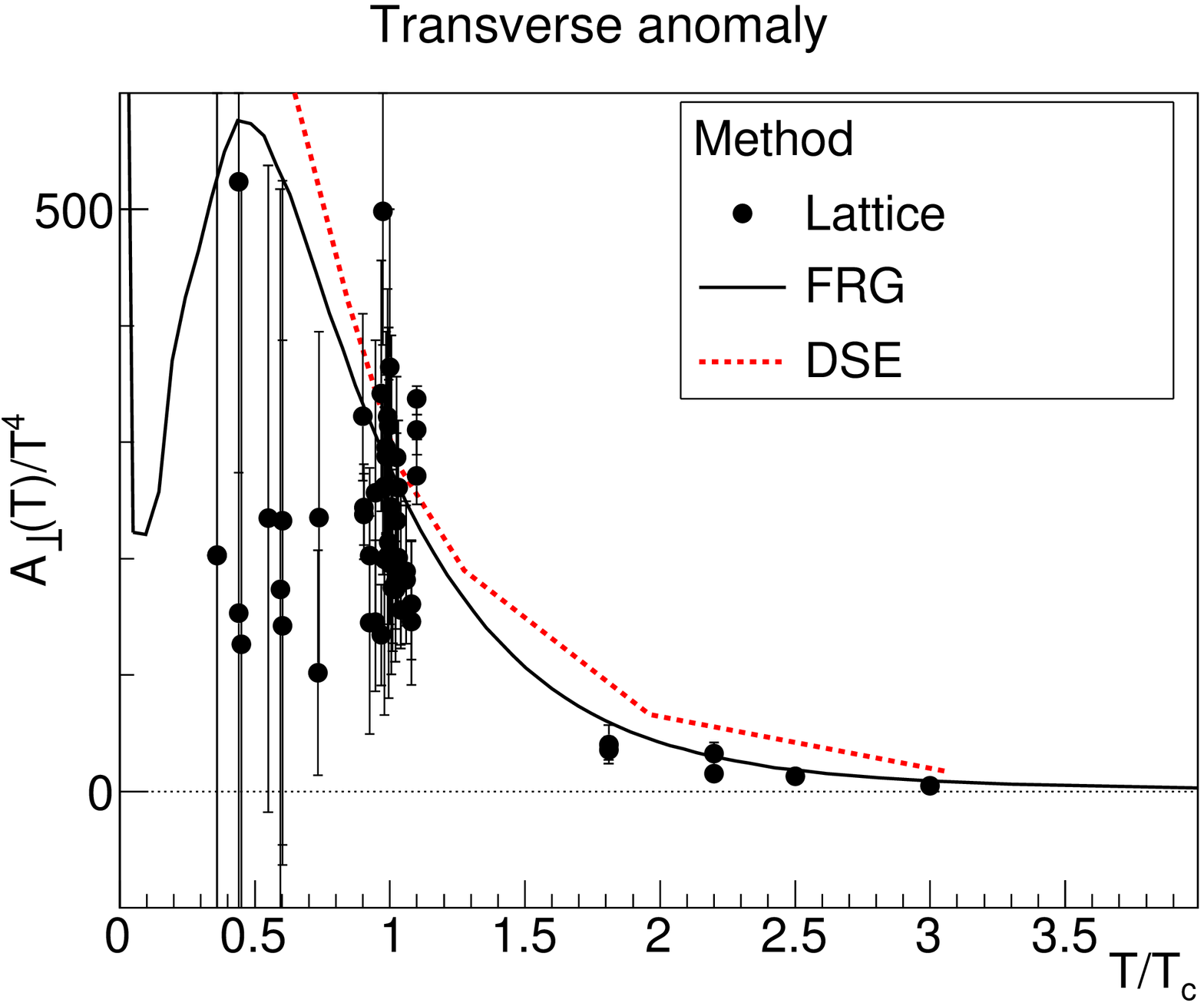}}{
                          \caption{Transverse anomaly $A_\bot(T)$ as a
                            function of the temperature.  \myhfill}
				\label{fig:at}
			}
			\ffigbox{\includegraphics[width=\columnwidth]{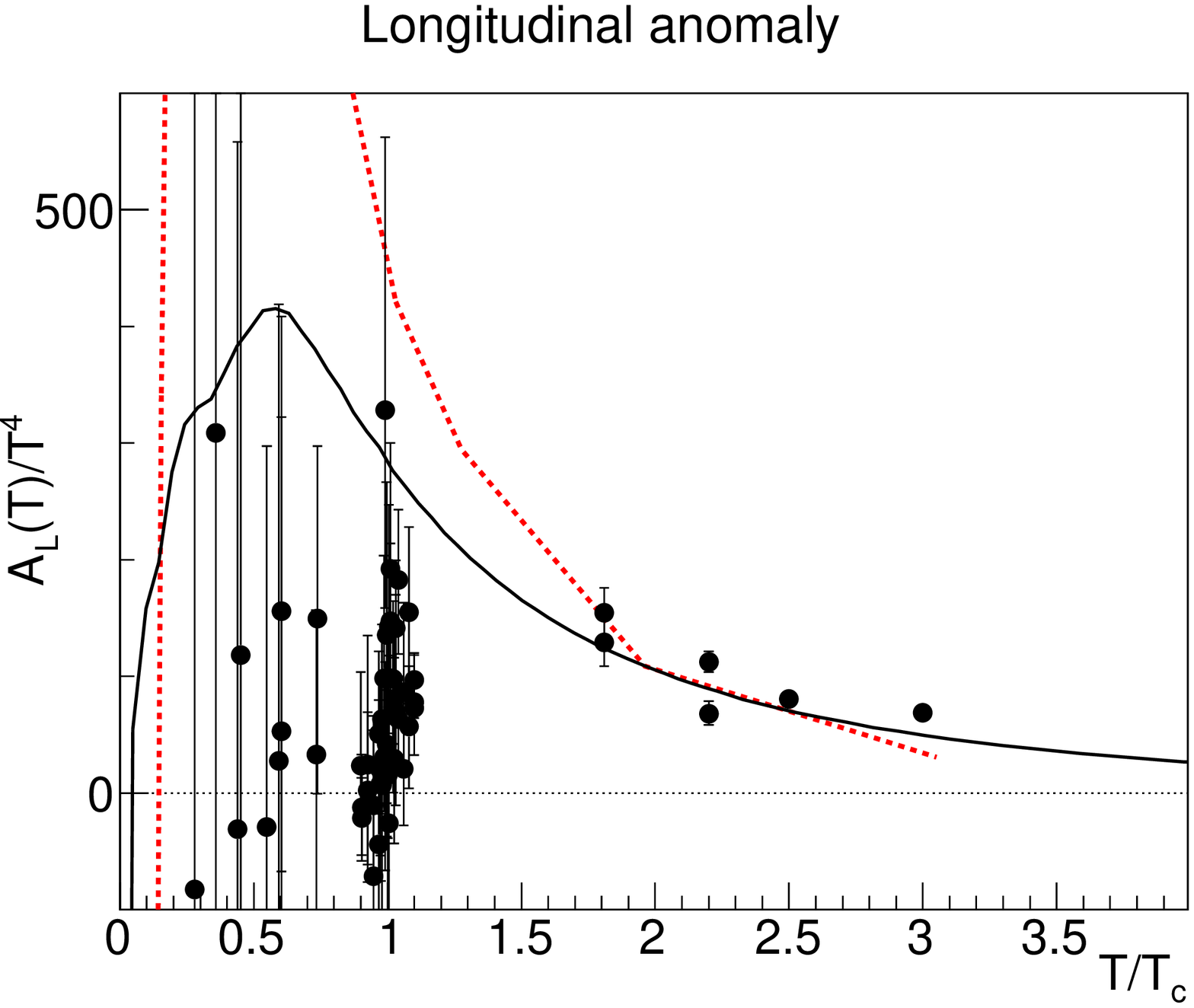}}{
                          \caption{Longitudinal anomaly $A_L(T)$ as a
                            function of temperature.\myhfill}
				\label{fig:al}
			} 
		\end{subfloatrow}
	}{
		\caption{
		The anomaly obtained from the Tan contact term from lattice, FRG \& DSE. 
		}
		\label{fig:AYangMills}
	}
\end{figure*}

While the Tan contact term in solid state physics and ultracold atoms
encodes the thermodynamics, it is in itself not yet equivalent to a
thermodynamic potential. However, it is linked to the thermodynamic
anomaly $A(T)$, see e.g.\ \cite{Enss:2019dv},
\begin{align} A(T)=\beta(g(T))C(T)
\label{eq:Tan-Anomaly}  \end{align}
wherein $\beta(g)$ is the $\beta$-function and $g(T)$ the
temperature-dependent running coupling evaluated at the
temperature. An analogous derivation in Yang-Mills theory faces
several intricacies. First of all this concerns the unphysical
nature of gluon fields in comparison to that in solid state and
ultracold atomic systems. This leads us to negative norm states in the
Fock space as well as the occurrence of ghost fields. Accordingly, a
Yang-Mills analogue of the relation \eqref{eq:Tan-Anomaly} will involve
$C_\bot, C_L$ and $C_\textrm{ghost}$ and the respective
$\beta$-functions $\beta_\bot, \beta_L, \beta_\textrm{ghost}$ as well
additional normalisation factors. The latter differ in the strongly-correlated
low temperature regime with $T\lesssim T_c$. Being short of the full
resolution of the different ingredients of the Yang-Mills relation we
here discuss the chromomagnetic and chromoelectric parts of this
relation. They are given by
\begin{align}\label{eq:Tan-AnomalyYM} 
  A_{\bot/L}(T)\approx \beta(g(T))C_{\bot/L}(T)\,,
\end{align}
where we will use the same $\beta$-function for chromomagnetic and
chromoelectric parts and take the normalisation factors to unity.

Note that the left-hand side of \eqref{eq:Tan-AnomalyYM} are related
to an observable, the thermodynamic or trace anomaly in Yang-Mills
theory. Thus, scheme-dependencies on the right-hand side need to
cancel, implying that the Tan contact term is scheme-dependent. In
addition, the miniMOM or Taylor scheme
\cite{vonSmekal:2009ae,Boucaud:2008gn} employed in the calculation of
the gluon propagators yields a multi-valued $\beta$-function and its
precise determination in lattice calculations requires high
statistics. While the former can be remedied by using the
temperature-dependent correct branch, the latter precludes us yet from
a full determination within each method separately. Also, as will be
seen, the Tan contact term needs to be determined at much higher
precision in the low-temperature domain.

However, as a proof-of-principle, we will use here an analytic,
temperature-independent coupling motivated by analytic perturbation
theory,
\be
\alpha(p)=\pi\frac{\ln\frac{\Lambda_c^2}{\Lambda_\text{YM}^2}}{
  \ln\frac{\Lambda_c^2+p^2}{\Lambda_\text{YM}^2}}\nn
\ee \no
taking $\Lambda_c^2=1.21$ GeV$^2$ for the cutoff momentum and
$\Lambda^2_\text{YM}=0.81$ GeV$^2$ for the scale. Note that this will
necessarily upset the overall scale of the result, as we do not use
matched schemes.

The results are shown in figure \ref{fig:AYangMills}. The lattice
results, albeit with large errors, are consistent with the temperature
dependence of the anomaly, showing a peak around the phase transition,
and a slow decrease towards large temperatures. At low temperatures,
where already the Tan contact term is compatible with zero within the
errors, so is necessarily the anomaly. Moreover, we deduce from figure
\ref{fig:AYangMills} that the overall normalisation
of \eqref{eq:Tan-AnomalyYM} is non-trivial as the trace anomaly
$A_\textrm{YM}$ in Yang-Mills obeys $A_\textrm{YM}\lesssim 3$, see
e.g.\ \cite{Borsanyi:2012ve}, while $A_{\bot/L} \lesssim 10^3$. Both
functional results show a quite similar behavior at high temperatures,
but tend to have the peak at far too low temperatures. This is likely
partly because this temperature regime is in the deep infrared, where
the $\beta$-function is not dominated by its perturbative
behavior. Here, a determination of the $\beta$-function in a
consistent scheme would likely cure these problems.

Nonetheless, the anomaly shows qualitatively the expected behavior,
indicating that the Tan contact term may indeed be a suitable approach
to obtain thermodynamic information from propagators.

\section{Conclusions}\label{s:con}

We have extracted for the first time the Tan contact term for
Yang-Mills theory from the gluon propagator. We see that known
thermodynamic features, the phase transition, the asymptotic
Stefan-Boltzmann behavior, and the strongly-interacting liquid
behavior imprint themselves qualitatively in the Tan contact term. We
also see that the various effects distribute themselves among the
transverse and longitudinal degrees of freedom differently. While the
strong-interaction regime above the phase transition seems to be
encoded in the chromomagnetic sector, the critical and bulk behavior
seems to be carried by the chromoelectric sector. This agrees with
observations in the infrared \cite{Maas:2011ez}. It has also been
shown that it is, in principle, possible to use the Tan contact term
to determine the anomaly and thus thermodynamic bulk properties.

The obvious steps to be taken from here are to improve statistics and
systematics on the lattice and to compare to further results from
other sources, e.\ g.\ hard-thermal loop calculations or results from
dimensionally-reduced calculations
\cite{Blaizot:2001nr,Hietanen:2008tv}. Another issue are contributions
from the ghost, which at first sight seems to be inert to temperature
\cite{Maas:2011ez,Huber:2013yqa,Fister:2014bpa,Cyrol:2017qkl}. For a
reconstruction of the thermodynamic potential in full it is required
to find the correct normalisation, a suitable scheme and sufficient
precision to determine the anomaly. Finally, an extension to finite
density is of high interest. Here, also QCD-like theories without sign
problem, e.\ g.\ 2-color QCD or G$_2$-QCD, could be interesting
testing grounds.


\begin{acknowledgments}

  We are grateful to T.~Enss and N.~Wink for discussions and a critical
  reading of the manuscript. We thank A.~K.~Cyrol for many discussion
  and a collaboration in an early phase of the project. O.~Hajizadeh
  was supported by the FWF Doctoral Program W1203 "Hadrons in vacuum,
  nuclei and stars" and by the mobility program of the Science Faculty
  of the University of Graz.  M.~Huber was supported by the FWF under
  Contract No. P27380-N27. The work is supported by EMMI, the BMBF
  grant 05P18VHFCA, and is part of and supported by the DFG
  Collaborative Research Centre "SFB 1225 (ISOQUANT)" as well as by
  Deutsche Forschungsgemeinschaft (DFG) under Germany’s Excellence
  Strategy EXC-2181/1 - 390900948 (the Heidelberg Excellence Cluster
  STRUCTURES). O.~Hajizadeh, M.~Huber, and A.~Maas are grateful for
  the warm hospitality at the University of Heidelberg during theirs
  visits there.

\end{acknowledgments}


\appendix

\section{Propagators from Dyson-Schwinger equations}
\label{sec:DSEs}
\begin{figure*}
	\ffigbox{
		\begin{subfloatrow}
			\ffigbox{\includegraphics[width=\columnwidth]{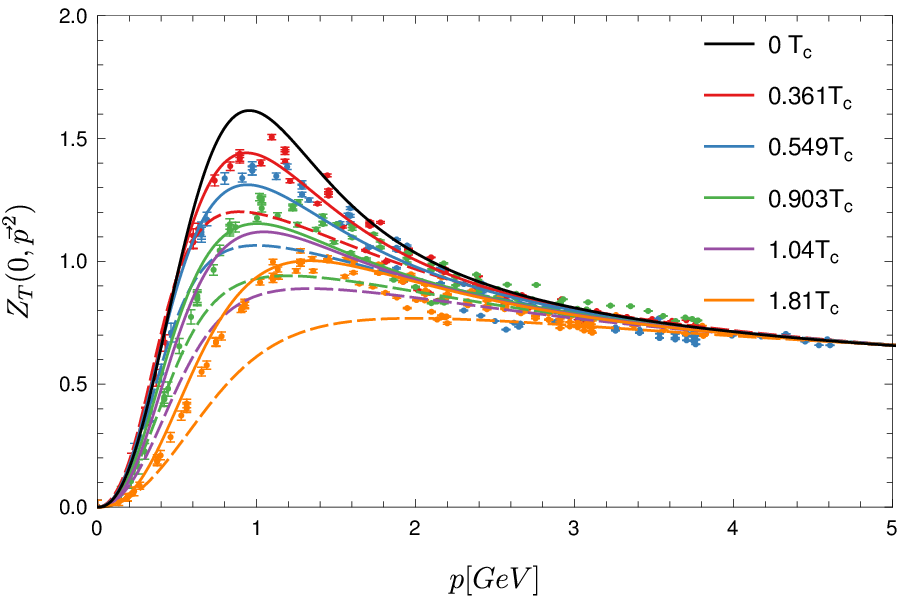}}{
				\caption{Chromo-magnetic gluon propagator
   dressings for selected temperatures.\myhfill}
				\label{fig:pZT}
			}
			\ffigbox{\includegraphics[width=\columnwidth]{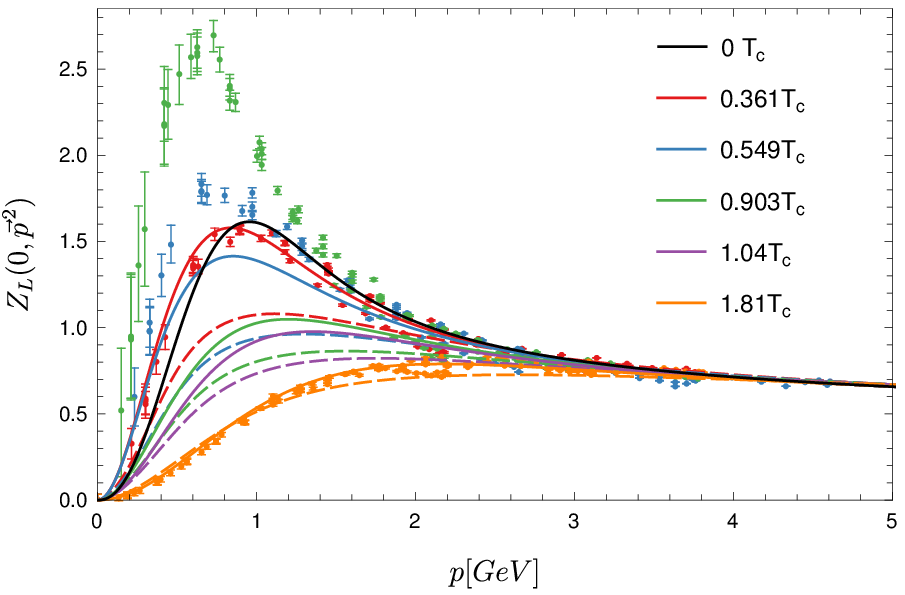}}{
				\caption{Chromo-electric gluon propagator
   dressings for selected temperatures.\myhfill}
				\label{fig:pZL}
			} 
		\end{subfloatrow}
	}{
          \caption{ Gluon propagator dressings for selected
            temperatures from the FRG (continuous), DSEs (dashed) and
            lattice simulations \cite{Fischer:2010fx,Maas:2011ez}. For
            reference the vacuum result from the FRG is shown in
            black.  }
		\label{fig:dressings_DSE}
	}
\end{figure*}

The truncation used for the calculation of the propagators from their
DSEs is described in the following.  The equations that were solved
are the ones for the ghost and gluon propagators truncated to one-loop
without tadpoles.  The only remaining higher $n$-point functions are
the ghost-gluon and three-gluon vertices.  The former is taken as
bare, which is within the context of the present work sufficient since
we are only interested in the high-momentum behavior.  The deviation
from a bare vertex is known to be a bump around $1\,\text{GeV}$ which
falls off quickly \cite{Fister:2014bpa,Huber:2016xbs,Cyrol:2017qkl}.

The three-gluon vertex plays a crucial role for the gluon propagator.
It is not only quantitatively relevant, but also the existence of a
solution for the gluon propagator depends strongly on its properties.
Here, the following model adapted from Ref.~\cite{Huber:2017txg} was
used for dressing the tree-level tensor,
\begin{widetext} 
  \begin{align}
    &C^{AAA}(p_0, q_0,\vec{p}, \vec{q})=
      \frac{G(\overline{p}^2)}{Z_T(\overline{p}^2)}
      \frac{\overline{p}^2}{\overline{p}^2+\Lambda_s^2}
      \Bigg(-G(\overline{p}^2)^3  \frac{\Lambda^2_\text{3g}}{
      \Lambda^2_\text{3g}+p^2}\frac{\Lambda^2_\text{3g}}{
      \Lambda^2_\text{3g}+q^2}\frac{\Lambda^2_\text{3g}}{
      \Lambda^2_\text{3g}+(p+q)^2}
      +\frac{G(\overline{p}^2)}{Z_T(\overline{p}^2)}\frac{
      \overline{p}^2}{\overline{p}^2+\Lambda_s^2}\Bigg).
\end{align}
\end{widetext}
The momentum $\overline{p}^2$ is $(p^2+q^2+(p+q)^2)/2$ and $p$ and $q$
are four-momenta.  $G$ and $Z_T$ are the ghost and the transverse gluon
dressing functions, respectively.  The first term in the parentheses
determines the IR behavior of the vertex, the second the UV behavior.
The term in front of the parentheses accounts for missing perturbative
higher loop contributions relevant for the resummed one-loop behavior
\cite{vonSmekal:1997vx,Huber:2012kd,Huber:2018ned}. The model contains
two scales which are fixed as
$\Lambda_s=\Lambda_\text{3g}=0.741\,\text{GeV}$.

The integral kernels for the Dyson-Schwinger equations can be found,
e.g., in Ref.~\cite{Maas:2005rf}.  Here they were derived with
\textit{DoFun} \cite{Alkofer:2008nt, Huber:2011qr, Huber:2019dkb}, and
the equations were solved with \textit{CrasyDSE} \cite{Huber:2012cd}.
Quadratic divergences in the gluon propagator DSE were renormalized
via second renormalization conditions chosen as the value of the
propagators at zero momentum \cite{Collins:1984xc, Meyers:2014iwa,
  Huber:2018ned}. For the employed truncation this leaves some
ambiguity how to select these conditions, but at the scales of
relevance here it is expected that such effects are subleading.

The overall scale was set for the lowest calculated temperature by
matching the UV tail to FRG results.  The relative scales for the
other temperatures were set by matching the perturbative couplings.

The resulting dressing functions for the gluon propagators are shown
in Fig.~\ref{fig:dressings_DSE} for selected temperatures. Clearly,
the present truncation cannot capture the IR behavior but reproduces
the momentum and temperature dependencies qualitatively. Both, in the
FRG and the DSE computation of the propagators the computation of the
non-trivial $A_0$-background has not been taken into account. This
background $A_0\neq 0$ is the equation of motion and is directly
linked to the vanishing of the Polyakov loop in the confining phase,
\cite{Braun:2007bx, Marhauser:2008fz, Braun:2010cy, Fister:2013bh,
  Herbst:2015ona}, for perturbative computations within the background
see \cite{Reinosa:2016iml}.

In
\cite{Cyrol:2017qkl} it has been argued that this should lead to
deviations of the chromo-electric propagators in functional approaches
from the chromo-electric lattice propagators (as they are computed on
different a background) for temperatures with
\begin{align}\label{eq:T-deviations} 
  0.5\, T_c \lesssim T\lesssim 1.3\, T_c\,.
\end{align}
Indeed this expectation is confirmed by the data, see figure
  \ref{fig:pZL}.

\bibliographystyle{bibstyle}
\bibliography{bib}


\end{document}